\documentclass[pdflatex,sn-mathphys-num]{sn-jnl}

\usepackage{graphicx}%
\usepackage{multirow}%
\usepackage{amsmath,amssymb,amsfonts}%
\usepackage{amsthm}%
\usepackage{mathrsfs}%
\usepackage[title]{appendix}%
\usepackage{xcolor}%
\usepackage{textcomp}%
\usepackage{manyfoot}%
\usepackage{booktabs}%
\usepackage{algorithm}%
\usepackage{algorithmicx}%
\usepackage{algpseudocode}%
\usepackage{listings}%
\usepackage{bm}
\usepackage{ulem}

\theoremstyle{thmstyleone}%
\theoremstyle{thmstyletwo}%

\theoremstyle{thmstylethree}%

\begin{document}

\title[Article Title]{
Observation of Strong Electron Correlation in Planetary Atomic Structure
}



\author[1,2,7]{\fnm{Xinglong}\sur{Yu}}
\equalcont{These authors contributed equally to this work.}
\author[3]{\fnm{Yongyan}\sur{Han}}
\equalcont{These authors contributed equally to this work.}
\author*[2]{\fnm{Zhenjie}\sur{Shen} }\email{shenzhj2@shanghaitech.edu.cn}
\author[3]{\fnm{Yong-Kang}\sur{Fang}}
\author[1,7]{\fnm{Shushu}\sur{Ruan}}
\author[1,7]{\fnm{Jie}\sur{Liu}}
\author[4,2]{\fnm{Zhixian}\sur{Wu}}
\author[2]{\fnm{Xincheng}\sur{Wang}}
\author[2]{\fnm{Ahai}\sur{Chen}}
\author[6]{\fnm{Wei-Chao}\sur{Jiang}}
\author[2,5]{\fnm{Kiyoshi}\sur{Ueda}}
\author*[3,8]{\fnm{Liang-You}\sur{Peng}}\email{liangyou.peng@pku.edu.cn}
\author*[2,4,1]{\fnm{Yuhai}\sur{Jiang}}\email{jiangyh3@shanghaitech.edu.cn}

\affil[1]{\orgdiv{Shanghai Advanced Research Institute}, \orgname{ Chinese Academy of Sciences}, \orgaddress{\street{Pudong}, \city{Shanghai}, \postcode{201210}, \country{China}}}
 
\affil[2]{\orgdiv{Center for Transformative Science and School of Physical Science and Technology}, \orgname{ShanghaiTech University}, \orgaddress{\street{Pudong}, \city{Shanghai}, \postcode{201210}, \country{China}}}


\affil[3]{\orgdiv{State Key Laboratory for Mesoscopic Physics and Frontiers Science Center for Nano-optoelectronics, School of Physics}, \orgname{Peking University}, \orgaddress{ \city{Beijing}, \postcode{100871}, \country{China}}}

\affil[4]{\orgdiv{School of Physics}, \orgname{Henan Normal University}, \orgaddress{ \city{Xinxiang}, \postcode{453007}, \state{Henan}, \country{China}}}
 
\affil[5]{\orgdiv{Department of Chemistry}, \orgname{Tohoku University}, \orgaddress{ \city{Sendai}, \postcode{980-8578}, \country{Japan}}}

\affil[6]{\orgdiv{Institute of Quantum Precision Measurement, College of Physics and Optoelectronic Engineering}, \orgname{Shenzhen University}, \orgaddress{\city{Shenzhen}, \postcode{518060}, \country{China}}}
  
\affil[7]{\orgdiv{School of Future Technology}, \orgname{University of Chinese Academy of Sciences}, \orgaddress{\city{Beijing}, \postcode{100049}, \country{China}}}

\affil[8]{\orgdiv{Collaborative Innovation Center of Extreme Optics}, \orgname{Shanxi University}, \orgaddress{\city{Taiyuan}, \postcode{030006}, \country{China}}}


\abstract{
Unravelling two-electron correlation is a long-standing challenge at the heart of few-body quantum physics, underlying correlated phenomena across atomic, molecular and condensed-matter science. In prototypical three-body Coulomb systems, such strong correlation in doubly excited states (DESs) of planetary atomic systems leaves distinct signatures in nonsequential above-threshold double ionization (NS-ATDI) driven by coherent laser fields, yet such targeted study has long remained elusive. Here we present kinematically complete measurements of multi-photon double ionization in cold strontium atoms. Our results reveal a dominant NS-ATDI channel exhibiting well-defined band structures that encode pronounced energy and angular correlations between the two emitted electrons. Autoionization spectra confirm the presence of DESs as transition states that effectively promote the NS-ATDI process. These observations provide direct evidence that both electrons are synchronously excited and ionized via resonant high-lying DES transitions, meaning the structure-linked two-electron correlation of DES is preserved and propagated in the laser-driven time-dependent three-body system. Our work transcends the traditional paradigm of multi-photon double ionization, and fundamentally reshapes the core understanding of intrinsic electron correlation governing many-body systems in nature.
}



\maketitle

Electron correlation in Coulombic systems forms the cornerstone of many-body physics and governs the electronic energies and structures of complex microscopic and macroscopic systems. It is especially critical for strongly correlated materials including high-temperature superconductors~\cite{science.235.4793.1196} and colossal magneto resistance in transition-metal oxides~\cite{RevModPhys.70.1039}. Studies on this topic often center on a prototypical platform provided by a helium or helium-like atom, where key phenomena including Fano resonance of doubly excited states (DESs) occur~\cite{science.1238396, science.1234407, Nature516_2014, Kotur2016, science.aah5188, science.aah6972, PhysRevLett.123.163201, Zhang2025}. These are the simplest three-body systems in nature, but complex enough to challenge the most advanced calculations~\cite{RevModPhys.72.497}. However, understanding the dynamics behind correlated electrons has been far from sufficient up to now.

DESs exhibit strong electron correlations characterized by interdependent motion of two electrons in position and momentum space~\cite{PhysRev.124.1866, PhysRevLett.10.516}. When two electrons in high-lying DESs orbit the nucleus in distinct paths, analogous to planets orbiting a star, the system constitutes a so-called planetary atom~\cite{10.1098/rspa.1977.0035} —a strongly correlated three-body system that has attracted increasing interest for decades~\cite{PhysRevLett.65.1965, PhysRevLett.64.274, PhysRevLett.68.21, RevModPhys.72.497, PhysRevA.101.013414, PhysRevA.104.012812}.
Ion-yield and absorption spectroscopy have served as standard experimental tools to investigate planetary atoms~\cite{PhysRevA.78.021401,PhysRevA.104.012812, PhysRevA.108.012816}, but such studies have been largely limited to measuring the energies of DESs. Direct access to the underlying structural dynamics has remained elusive, even though classical Kepler planetary motions in the eZe and Zee configurations have been predicted to exhibit angular stability~\cite{RevModPhys.72.497,PhysRevA.78.021401,PhysRevA.82.033422}. In these simplified one-dimensional pictures, the eZe configuration corresponds to two electrons on opposite sides of the nucleus, whereas both electrons in the Zee configuration are on the same side. Remarkably, the angular correlation of the electrons is preserved from high-lying DESs into the continuum above the double ionization~(DI) threshold, even across the quantum chaotic regime~\cite{PhysRevA.78.021401, PhysRevA.82.033422, PhysRevA.84.023402, Jiangyh2006}.

DI serves as a sensitive probe of such correlated behavior, especially nonsequential DI, because the energy and momentum spectra of the two outgoing correlated electrons carry rich information that reveals the underlying physical mechanism.
As an outstanding benchmark, single-photon DI studies have uncovered the canonical knock-out and shake-off ionization mechanisms. 
Investigations have been extended to two-photon~\cite{PhysRevLett.98.203001, PhysRevLett.101.073003, Kurka_2009, PhysRevLett.115.153002, PhysRevLett.129.183204}
, multi-photon~\cite{Ni_2011, PhysRevA.97.031405}, and strong-field induced~\cite{PhysRevLett.73.1227, Weber2000,PhysRevLett.124.043203} DIs, revealing rich intrinsic ionization dynamics that remain active research frontiers. In the multi-photon strong perturbation regime, the above-threshold ionization phenomenon for two-electron emission—known as nonsequential above-threshold double ionization (NS-ATDI)—gives rise to highly correlated two-electron wave packets, in which both electrons absorb photons and share the excess energy in units of one photon. Using numerical solutions to the full-dimensional two-electron time-dependent Schr\"{o}dinger equation~(TDSE), Parker {\it {et al.}}~\cite{JSParker_2001} calculated ATDI for helium in high-coherence, high-brightness XUV laser fields and found a hint of nonsequential signals in the coexistence of primary sequential DI processes. There have been experimental reports on ATDI of rare gas atoms~\cite{ArATDI, PhysRevA.98.043405} and molecules~\cite{c2h2ATDI} induced by 400~nm lasers. To date, all experimental and theoretical studies indicate that sequential processes overwhelmingly dominate in ATDI.

Once high-lying DESs participate in NS-ATDI, a complex time-dependent three-body problem emerges, as the two-electron dynamics within the planetary structure evolves in the time-dependent laser field.
The strong radial and angular correlations encoded in transition DESs~\cite{science303.813, PhysRevLett.110.023002} can be accessed by detecting the emitted electron pairs, yielding distinct and revealing signatures in NS-ATDI.
However, NS-ATDI studies involving DESs as intermediate transition states remain poorly developed because of substantial methodological challenges. Theoretically, single-photon perturbation theory and simple mean-field approaches are insufficient to describe this physical process. Furthermore, owing to multi-photon absorption, the coherence of the laser field is imprinted onto the highly correlated transition and emission processes, posing a great challenge for achieving a clear physical picture.

\begin{figure}[h]
\centering
\includegraphics[width=1.0\textwidth]{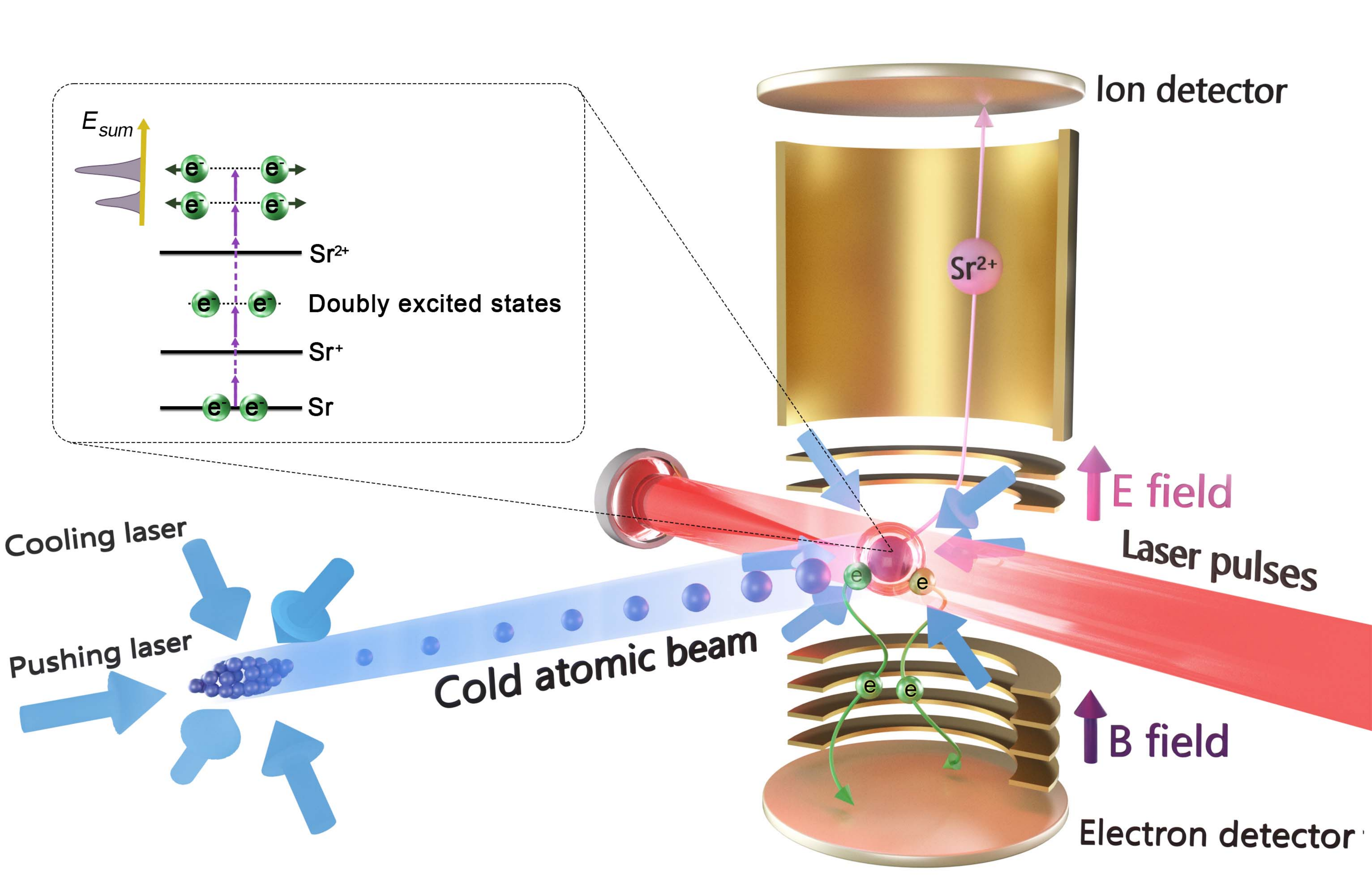}
\caption{\textbf{Schematic of the experimental setup}. The continuum cold atomic beam is delivered to the reaction chamber by a red-shifted 461~nm pushing laser and further ionized by femtosecond lasers. The photoelectrons and recoil {Sr$^{2+}$} ion are guided to detectors at opposite ends by homogeneous magnetic and electric fields. The inset illustrates that a pair of electrons is ionized with correlated kinetic energies and emitted back-to-back by the proposed {NS-ATDI} mechanism via intermediate DESs.}\label{fig1}
\end{figure}

In this work, we use strontium atom—a helium-like system—as an ideal platform to investigate two-electron correlations. In this system, a large number of DESs can be efficiently populated via multi-photon excitation with optical lasers~\cite{PhysRevA.61.040502, PhysRevA.97.013429, PhysRevA.104.012812, PhysRevA.108.012816, svsd-9mj3}, owing to much larger excitation cross sections than those of helium with one-photon absorption.
Specifically, we perform the first kinematically complete study on ATDI by exploiting a state-of-the-art magneto-optical trap reaction microscope~(MOTReMi)~\cite{SrMOTREMI, SrSI_2024, PhysRevLett.134.123204} and numerically solving the full-dimensional two-electron TDSE~\cite{ PhysRevLett.124.043203,PhysRevLett.115.153002,PENG20151}. A schematic diagram of Sr-MOTReMi is shown in Fig.~\ref{fig1}. The cold strontium atoms created by laser cooling technologies were doubly ionized by femtosecond lasers. The momenta of Sr$^{2+}$ ion and two photoelectrons in three dimensions were detected in coincidence with extremely high resolutions. The joint energy spectra (JES) from DI of the ground state {$5s^2(^1S_0)$} {show} significantly dominant {NS-ATDI features} where two electrons present band structures and strong angular correlations as equally sharing excess energy. As intermediate transition states, DESs relay the electron correlations~(see in the inset of Fig.~\ref{fig1}), thereby enabling these distinct NS-ATDI features in the laser-driven time-dependent three-body system. 
 
\section*{ATDIs at 800~nm}\label{sec1}

\begin{figure}[h]
\centering
\includegraphics[width=1.0\textwidth]{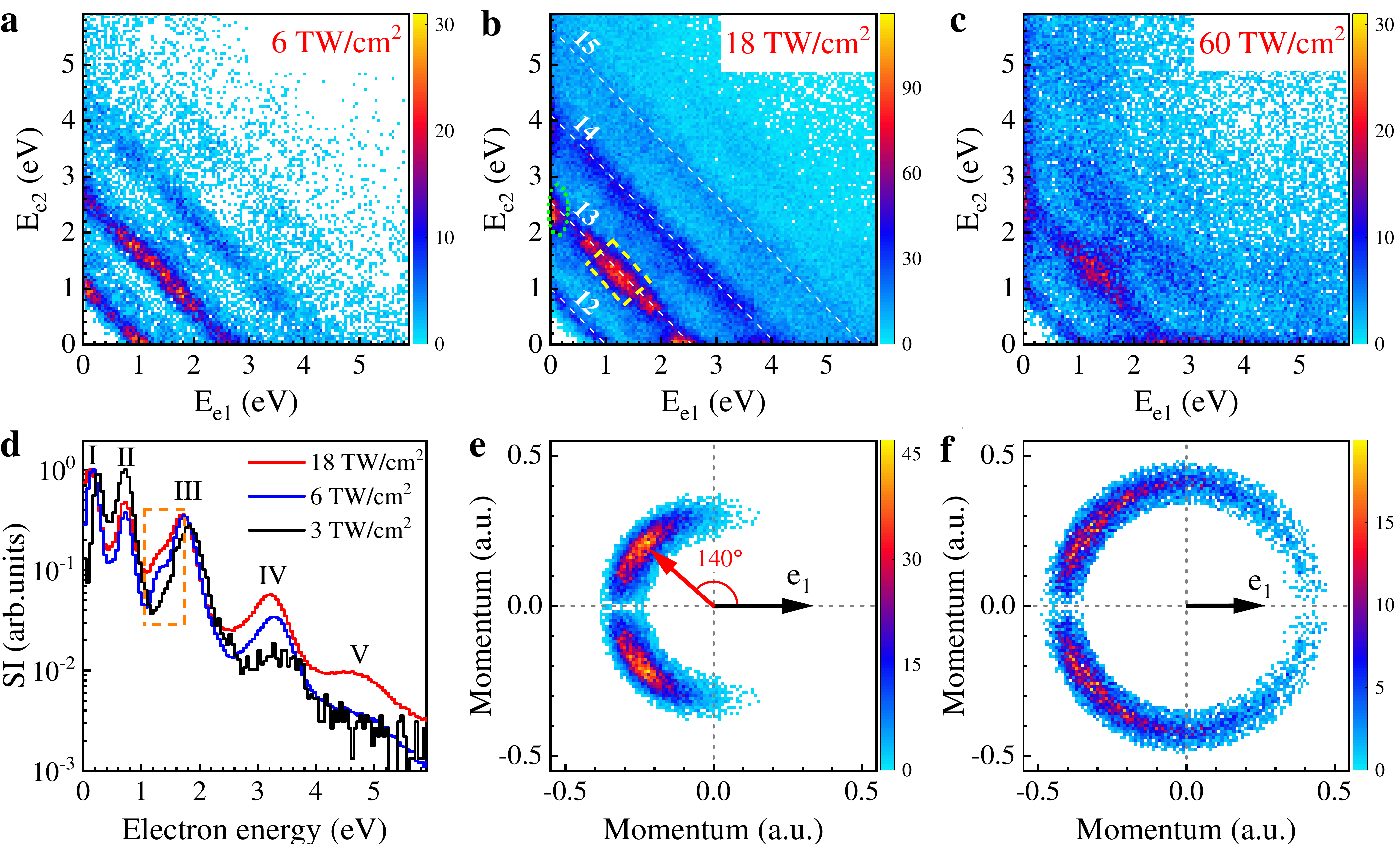}
\caption{\textbf{Electron energy spectra and correlated angular distributions (CADs)}. \textbf{a--c}, electron-electron JES at $I=$ 6 TW/cm$^2$, 18 TW/cm$^2$, and 60 TW/cm$^2$. In \textbf{b}, the white dashed line presents the zeroth to third ATDI where kinetic energy-sum of two electrons are constant and 12-15 indicate the number of photons absorbed. \textbf{d}, electron energy spectra for SI at $I=$ 3 TW/cm$^2$, 6 TW/cm$^2$, and 18 TW/cm$^2$. The peaks \uppercase\expandafter{\romannumeral1}, \uppercase\expandafter{\romannumeral3}, \uppercase\expandafter{\romannumeral4} and \uppercase\expandafter{\romannumeral2} correspond to electrons ionized from $5s^2\ ^1S_0$ with absorption of four-photon, the five-photon, six-photon and from $5s5p\ ^3P_2$ with three-photon absorption, respectively. The spectra are normalized to maxima of the peak \uppercase\expandafter{\romannumeral1}. \textbf{e}, CADs of the second electron with $E_{e2}=1.0-1.5$~eV as the first electron with $E_{e1}=1.0-1.5$~eV emitted along the positive x-axis, corresponding to nonsequential ATDI electrons marked within the yellow dashed box in \textbf{b}.  \textbf{f}, same as \textbf{e} but for sequential ATDI with $E_{e2} = 2.2 - 2.6$~eV and $E_{e1} = 0 - 0.2$~eV, corresponding to electrons marked within the green dashed circle in \textbf{b}. }\label{fig2}
\end{figure}

In Fig.~\ref{fig2}a--c, we present the JES for DI of {$5s^2(^1S_0)$} driven by a linearly polarized 800~nm laser at various laser intensities, and electron-electron joint momentum spectra (JMS) can be seen in the Supplementary Fig.~S1. These results clearly demonstrate the multi-photon absorption in a strong perturbative regime. The total energy for each ATDI structure is given by $E_{\rm{sum}} = E_{e1} + E_{e2} = n \omega - I_{p1} - I_{p2} - 2 U_p$, where $ \omega$ is a photon energy, $E_{e1}$ and $E_{e2}$ represent the kinetic energies of the two photoelectrons , and $I_{p1}$= 5.7~eV/$I_{p2}$ = 11.0~eV denote the first/second ionization potential. The number of absorbed photons $n$ is estimated to be $n=12$-$15$. {$U_p$ $\approx$ 0.4~eV is the ponderomotive shift at a laser peak intensity $I$ = 6 TW/cm$^2$. For the sequential DI of $5s^2(^1S_0)$, four and eight-photon are required for the ionization processes Sr $\rightarrow$ Sr$^+$ and Sr$^+\rightarrow$ Sr$^{2+}$, respectively. For {the} nonsequential DI {of $5s^2(^1S_0)$}, at least} 11 photons are required. However, the 11-photon channel is effectively closed due to the shift induced by $U_p$. 

Distinct NS-ATDI features are observed in the JES. Taking the 13-photon channel in Fig.~\ref{fig2}b as an example, NS-ATDI shows a pronounced band pattern in the middle of the diagonal stripe, which is also characterized by circular arcs in the JMS (see Fig.~S1 in the Supplementary). Two emitted electrons share a total energy of $E_{\rm{sum}} = 13 \omega - I_{p1} - I_{p2} - 2 U_p \approx 2.7$~eV, i.e., the kinetic energy of one electron increases as the other one decreases or vice versa, keeping their total energy constant. In contrast, sequential ATDI appears as island structures, one of which is indicated in the green dashed circle in Fig.~\ref{fig2}b. In this case, 
the two sequentially ionized electrons have different kinetic energies, with $E_{e1} \approx0.1$~eV from the single ionization (SI) of Sr~( see also the peak \uppercase\expandafter{\romannumeral1} in Fig.~\ref{fig2}d) and $E_{e2} \approx2.4$~eV from the first order above-threshold ionization (ATI) of Sr$^+$. As the laser intensity increases, these features of NS-ATDI always prevail, although the stripes start to become broad because of the varying $U_p$ over the entire laser pulse. The continuous distribution of these band structures constitutes a definitive hallmark of nonsequential ATDI and reflects two-electron energy correlations induced by multi-photon absorption. Most notably, the nonsequential signals are remarkably prominent, even comparable to those of sequential processes. These findings are quite different from previous theoretical and experimental ATDI studies~\cite{JSParker_2001,ArATDI, PhysRevA.98.043405}, where sequential ATDIs always dominate absolutely. For sequential ATDI, two electrons are sequentially ionized from Sr and Sr$^+$ and their kinetic energies depend on the number of absorbed photons  and the ionization potentials I$_p$, resulting in the absence of energy correlations. 

Strong electron correlations are further manifested in the correlated angular distribution (CAD) of the two photoelectrons. For nonsequential electron pairs with nearly equal kinetic energies $E_{e1} = E_{e2} \approx 1.0-1.5$~eV located in the yellow dashed box in Fig.~\ref{fig2}b, the CAD of one electron is plotted in Fig.~\ref{fig2}e with the emission direction of the other fixed. The maximum ionization probability of the second electron appears at a relative emission angle $\theta_{12}=140^\circ \pm 10^\circ$, indicated by a red arrow. 
The emission pattern of the almost back-to-back might refer to the planetary atomic structure of the transition DESs, where the classical eZe configuration reflecting angular correlation survives significantly at relatively large angles between two electrons for the principal series~\cite{PhysRevA.78.021401, PhysRevA.82.033422}.  For the principal series of DESs, the angle $\theta_{12}$ can approach 180$^\circ$ as closely as possible when approaching the DI threshold~\cite{PhysRevA.78.021401,PhysRevA.82.033422}. The enormous density of DESs might lead to nontrivial mixing, so even qualitative analysis of the DESs can be difficult. Nevertheless, a detailed and complete understanding still challenges current many-body theories and needs further investigation. The CAD for two sequential electrons is also shown in Fig.~\ref{fig2}f. As expected, we observe almost isotropic distributions, i.e., one electron is emitted independently on the other, and no maximum in CAD is present. Note that progressively decreasing counts at relatively small angles mainly results from the decreasing detection efficiency when two electrons arrive at the detector with approaching time. It has been confirmed that the CAD profile for NS-ATDI in Fig.~\ref{fig2}e does not change after calibration with the data of sequential ATDI in Fig.~\ref{fig2}f. In addition, the data in the regions very close to {0$^\circ$ and} 180$^\circ$ are not physically meaningful due to the solid angles.

Considering the two-electron energy levels of strontium, at 800~nm, multi-photon resonant excitation to DESs should have been involved in the intermediate stage. Subsequently, the outermost two electrons in highly excited orbitals can be simultaneously ionized with nearly equal energy sharing and back-to-back emission. To verify the role of DESs, we plot the energy spectra of the SI in Fig.~\ref{fig2}d at various laser intensities, where electrons are collected in coincidence with Sr$^+$ in order to avoid contributions from DI. Obviously, the distribution of the peak \uppercase\expandafter{\romannumeral3}, i.e., the first ATI, gradually varies from being symmetric to being asymmetric as the laser intensity increases, indicating the onset of autoionization decay from DESs. Most importantly, the autoionization features within the dashed box coincide closely with the emergence of NS-ATDI: At $I=3 $ TW/cm$^2$, we find no DI signals, while the autoionization of DESs is also negligible in Fig.~\ref{fig2}d. As autoionization signals gradually present at $I= 6$ TW/cm$^2$ and 18 TW/cm$^2$, NS-ATDI structures accordingly become correspondingly prominent. However, it remains difficult to identify which specific DESs give rise to the spectral change near the first ATI because of the presence of a dense manifold of DESs. Nevertheless, for the first time, the above experimental observations present the scenario that dominating NS-ATDI features result from simultaneous emission of two electrons by coherent absorption of a couple of photons via one or a few DESs. The pronounced energy and angular correlations of two-electron pairs directly reflect the characteristic ionization dynamics of DESs.

\section*{ATDIs with different initial state and ellipticity}\label{secA2}

\begin{figure}[h]
\centering
\includegraphics[width=1.0\textwidth]{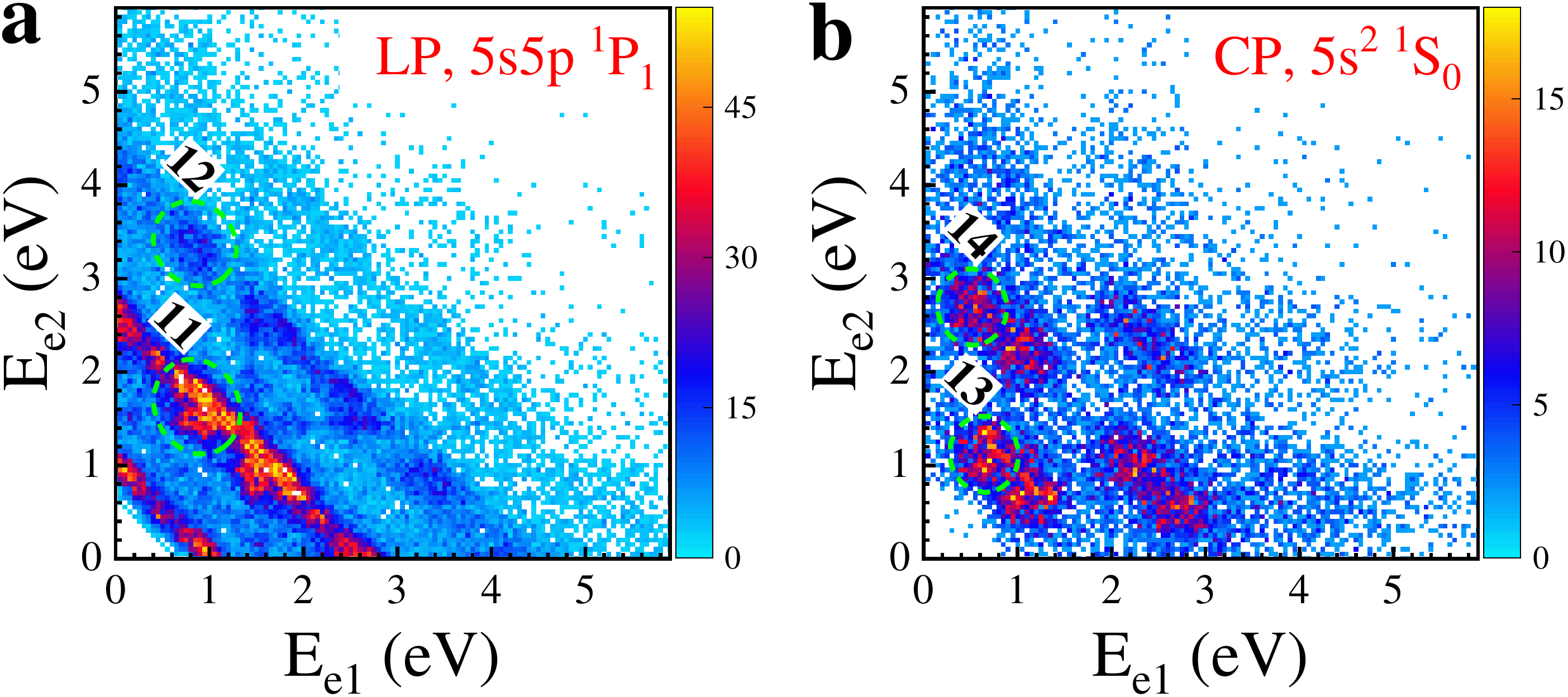}
\caption{\textbf{The initial-state- and ellipticity-dependent JES.} \textbf{a}, the JES from the singly excited state $5s5p\ ^1P_1$ by a linearly polarized (LP) pulse at $I$=6 TW/cm$^2$. \textbf{b}, the JES from the ground state but in a circularly polarized (CP) pulse at $I$=18 TW/cm$^2$. The corresponding numbers of photons required for ATDI are indicated in each channel.}\label{figure3}
\end{figure}


Furthermore, we point out that the pathways to intermediate DESs can be manipulated by changing initial states and the ellipticity of the laser pulse in Fig.~\ref{figure3}. Interestingly, the JES in Fig.~\ref{figure3}a presents visible and discrete island structures of sequential ATDIs, distinctly different from those in Fig.~\ref{fig2}a. Changing the initial state from $5s^2(^1S_0)$ to $5s5p(^1P_1)$ will modify the character of intermediate transition states, which strongly perturbs electron correlations and ultimately suppresses the NS-ATDI channel. Even this kind of distinguishability of initial electrons with different orbital quantum numbers favors the ionization of one electron first, consequentially resulting in the sequential ionization. This partly supports the present scenario that both electrons are ionized simultaneously via transition DESs, where they jump together via DESs ladders forming planetary atomic structure. Meanwhile, by using a circularly polarized laser to ionize the $5s^2(^1S_0)$ atoms, sequential ATDIs with clear ‘islands’ structures cover the JES, as shown in Fig.~\ref{figure3}b. This may be explained as follows: once a circularly polarized photon is absorbed, the quantum number of the total angular momentum prefers to increase by one, so that it is highly detrimental to the formation of DESs during the absorption of more than 12 photons. Consequently, the strong correlation is intercepted during the transition stages, and the electrons tend to be released one-by-one.

\section*{ATDIs at 400~nm}\label{sec3}

\begin{figure}[h]
\centering
\includegraphics[width=1.0\textwidth]{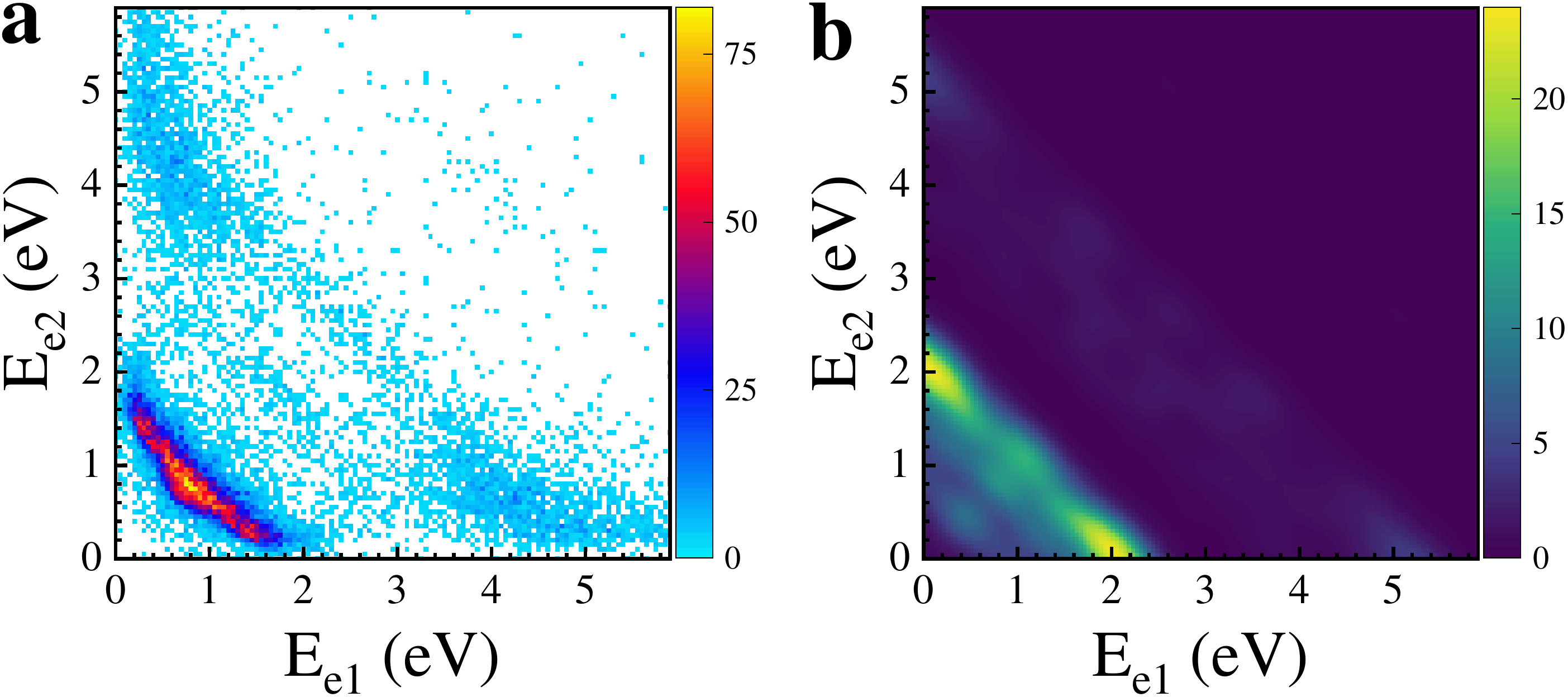}
\caption{\textbf{Same as Fig.~\ref{fig2}a--c but for linearly polarized 400~nm pulse at 18 TW/cm$^2$}. \textbf{a}  the experimental electron-electron JES; \textbf{b}  full-dimensional TDSE simulations.}\label{fig4}
\end{figure}

To the best of our knowledge, so far, there is no report on a rigorous calculation of multi-photon induced ATDI at 800~nm or longer wavelengths for heavy atoms with two outermost electrons. The extremely large computation demand prohibits the attempt under current theoretical methodologies. 
To further consolidate the above finding about the strong NS-ATDI, we carried out additional and independent measurements for a linearly polarized laser at 400~nm, whose results can be directly compared with our {\it {ab initio}} calculations. 

In this shorter wavelength region, the field-induced electron collision is further suppressed and potential pathways of resonant excitation are also simplified. The experimental JES together with the results of the two-electron TDSE calculation are shown in Fig.~\ref{fig4}, where the zero and first orders of ATDI are observed at $I$ = 18 TW/cm$^2$. We note that the nonsequential DI band in the experimental measurements remains the dominant structure located at $E_{e1} = E_{e2}\approx 0.8$~eV, whose overall features are reproduced by the theoretical calculation. The simultaneous excitation of {DESs} during the multi-photon absorption, as we discussed before, is very likely a universal correlation mechanism that induces the strong signal in the NS-ATDI channel.

\section*{Conclusions}\label{sec4}
Combining a state-of-the-art MOTReMi apparatus with full-dimensional two-electron TDSE numerical simulations, we have carried out kinematically complete investigations into multi-photon-induced ATDI in strontium atoms. Our experimental measurements unambiguously reveal prominently dominant NS-ATDI, with the emitted correlated electron pairs exhibiting pronounced energy and angular correlations that refer to the underlying atomic structure of doubly excited states (DESs). Complementary autoionization spectra further reveal distinct contributions originating from DES autoionization, in consistent with the presence of NS-ATDI. This finding delivers direct experimental evidence that  high-lying DESs play an indispensable role in mediating the strong electron-electron correlation throughout the ionization process. These remarkable, previously unreported nonsequential features stem from coherent excitation pathways: two tightly bound electrons simultaneously absorb coherent laser photons, and undergo stepwise excitation and ionization via intermediate DES ladders. This work marks a substantial advance in deciphering strong two-electron correlation within laser-driven time-dependent three-body Coulomb systems, and provides fundamental insights into electron correlation dynamics spanning both microscopic and macroscopic quantum phenomena.

\section*{Methods}\label{sec2}
\subsection*{Experimental methods}
A schematic diagram of the experimental {setup}, the Sr-MOTReMi~\cite{SrMOTREMI,SrSI_2024}, is shown in Fig.~\ref{fig1}. The hot strontium atoms evaporated in an oven are decelerated by a Zeeman slower laser beam and then precooled and trapped by two orthogonal pairs of retroreflected two-dimensional (2D) cooling laser beams. Subsequently, the 2D MOT is propelled to generate a cold atomic beam by a pushing laser beam and further cooled and trapped by a three-dimensional (3D) MOT configuration. Consequently, cold atomic targets for our setup can be prepared in the 2D MOT beam, molasses beam, and 3D MOT. The 2D MOT beam delivers the continuous cold atomic beam employed for photoionization measurements of Sr($5s^2\ ^1S_0$). When the six 3D cooling laser beams are turned on in the reaction chamber without the gradient magnetic field, the molasses target containing cold atoms in the Sr($5s5p\ ^1P_1$) state is prepared for measurements. The 3D MOT is used for the overlapping determination of laser focus and atomic beams. We used a Ti:sapphire laser pulses with 800~nm and 400~nm wavelengths and 50~fs pulse duration (FWHM) at intensities of 3-60 TW/cm$^2$ to produce ions and electrons, which are collected in a 4$\pi$ angle in a standard ReMi approach, see detailed configurations in Section 1 of Supplementary Materials. The momentum resolution for coincident Sr$^+$/Sr$^{2+}$ and electrons along the longitudinal direction of the spectrometer is achieved up to 0.12~a.u. and 0.02~a.u., respectively, and slightly lower in the transverse plane. 

\subsection*{Theoretical methods}

The DI of the Sr atom is simulated using state-of-the-art numerical methods for solving the TDSE~\cite{PhysRevLett.124.043203,PhysRevLett.115.153002,PENG20151}. Here, we assume that only two electrons are active and interact directly with the external laser field. The effects of inner-shell electrons and the nuclear core are incorporated through an angular-momentum-dependent pseudopotential
, for details see Section 3 in Supplementary Materials. The two-electron wave function is expanded in coupled spherical harmonics, while its radial components are discretized using the finite-element discrete variable representation (FE-DVR) method
. Both the maximum total angular momentum quantum number $L_{\max}$ and the maximum single-electron angular momentum quantum number $l_{\max}$ are set to 7. Nonuniform finite elements are employed, with finer element sizes near the core region and progressively larger elements at greater distances. Time propagation of the two-electron wave function is performed using the Lanczos method
. During propagation, the two electrons are confined within a spherical box of radius 250~a.u.. The wave function is periodically split into inner and outer components. The inner part is always propagated numerically under the full Hamiltonian, whereas the outer part is propagated analytically using the Volkov propagator
. At the end of the propagation, DI signals are extracted from both the inner and outer wave-function components.


\backmatter

\bmhead{Supplementary information}



Supplementary Materials contain details of the experiments, additional electron-electron joint momentum spectra, and the theoretical methods.

\bmhead{Acknowledgements}

This work is supported by the National Key Research and Development Program of China (Nos. 2022YFA1604302 and 2024YFA1612101), the National Natural Science Foundation of China (Nos. 12334011, 12234002, 12541505, 12204498, 12174284 and 12474346), and the Natural Science Foundation of Henan (Grant No. 252300421304). We also acknowledge support
from the Shanghai-XFEL beamline project (SBP) (Grant No.
31011505505885920161A2101001) and the computational resource of the HPC platform in ShanghaiTech University.

\section*{Declarations}

\begin{itemize}

\item Conflict of interest/Competing interests:

The authors declare no competing interests.

\item Data availability:

The data that support the plots within this article are available from the corresponding authors upon reasonable request.

\item Code availability:

The code that supports the plots within this article is available from the corresponding authors upon reasonable request.

\item Author contributions:

 Y.J. supervised the project. X.Y., Z.S., S.R., J.L., Z.W., X.W., and Y.J. performed the experiments. Y.H., Y.F., W.J., and L.Y.P. performed the simulations. X.Y., Z.S., and Y.J. analyzed the data. Y.J., Z.S., and X.Y. wrote the manuscript after in-depth discussions with K.U., A.C., and with important input from all other authors. 

\end{itemize}


\bibliography{sn-bibliography}

\end{document}



\title[Article Title]{
Supplementary Information for\\Observation of Strong Electron Correlation in Planetary Atomic Structure
}

\maketitle

\newpage
\tableofcontents

\newpage
\section{Details of the experiment}\label{secA1}
The Sr-MOTReMi setup employed in the experiments comprises two main components: a two-dimensional magneto-optical trap (2D MOT) that provides the cold atomic beam, and a reaction microscope for detecting the momentum vectors of recoil ions and electrons. More details have been reported in our previous work~\cite{SrMOTREMI}. 

Solid strontium is heated in an oven to $\sim$ 460 $^{\circ}$C to produce a thermal atomic beam into the 2D MOT chamber. The hot strontium atoms are decelerated by a Zeeman slower laser beam and subsequently precooled and captured by two pairs of orthogonally arranged and retroreflected 2D cooling laser beams to form the 2D MOT. A pushing beam with an intensity of 65 mW/cm$^2$ serves to transfer the 2D MOT to the reaction region, generating the cold atomic beam with the flux of $\sim$ 2 $\times$ 10$^7$ atoms/s. A weak uniform electric field of 1.7 V/cm accelerates the recoil ions and electrons, with a collinear homogeneous magnetic field ($\sim$ 4 G) applied to confine the electron trajectories. The acceleration lengths for recoil ions and electrons are 112.5 mm and 222.5 mm, respectively. The accelerated ions travel through a 430 mm drift region, whereas electrons undergo no drift section. These charged particles are collected by ion and electron detectors, each with a diameter of 80~mm, covering a 4$\pi$ solid angle. By measuring the time of flight (TOF) and the impact positions on the detectors, the three‑dimensional momentum vectors of the recoil ions and electrons can be reconstructed. The momentum resolution for coincident Sr$^+$/Sr$^{2+}$ and electrons along the longitudinal direction of the spectrometer reaches to 0.12~a.u. and 0.02~a.u., respectively, and is slightly worse in the transverse plane. To suppress potential false coincidence effects, the momentum conservation cutoff of $|p_{z,Sr^{+}} + p_{z,e}|< 0.5 $ a.u. is imposed for single ionization (SI), and $|p_{z,Sr^{2+}} + p_{z,e1}+p_{z,e2}|< 0.5 $ a.u. for double ionization (DI). Additionally, in the context of DI measurements, the probability of ionizing two atoms with the same laser pulse is below 2\%. 

We employed a Ti:sapphire laser system that delivers linearly polarized pulses centered at 800~nm with a pulse duration of 50~fs and a repetition rate of 1~kHz. The laser intensity can be continuously adjusted using a half-wave plate in conjunction with a linear polarizer. The pulse intensity is calibrated using the ponderomotive shift $U_p$ of electrons from SI of ground state atoms. At intensities below 6 TW/cm$^2$, the ground state atoms are directly ionized via the absorption of four photons, and the ac Stark shift is negligible. Hence, the measured electron energy shift is ascribed solely to $U_p$, from which the intensity is derived via $I = U_p \times 4 \omega^2$ (in atomic units). The laser intensities at higher power levels are determined by scaling this low intensity value according to the power ratio measured with a calibrated meter. Based on this procedure, the peak intensity of the 800~nm pulses used in our experiments is estimated to be 3-60 TW/cm$^2$. The linearly polarized 400~nm pulse is generated by doubling the frequency of 800~nm through a $\beta$-barium borate crystal, and the peak intensity is determined to be 18 TW/cm$^2$. Across the intensity range studied, the maximum electron counting rate remained below 0.1 electrons per laser pulse, with a consistent ion-to-electron ratio of approximately 1:3.

The cold atomic beam contains strontium atoms in both the ground state and the metastable state $5s5p\ ^3P_2$. The $5s5p\ ^3P_2$ state is a dark state that is generated and accumulates within the 2D MOT and is subsequently transported together with ground state atoms into the reaction chamber and exhibits a spatial distribution peaked away from the laser focus. The SI of $5s5p\ ^3P_2$ is easily achieved, whereas above-threshold ionization (ATI) and above-threshold double ionization (ATDI) involving the absorption of more photons are strongly suppressed. The molasses target is created by 3D cooling laser beams without gradient magnetic field, with $\sim$ 10\% of the atoms populated into the $5s5p\ ^1P_1$ state. In comparison, the $5s5p\ ^3P_2$ state generated in the reaction region by the 3D cooling beams accounts for only approximately 1/150000 of the population in the $5s5p\ ^1P_1$ state, rendering it negligible. Consequently, DI originating from the $5s5p\ ^3P_2$ state is not discussed in this work.

\section{Electron-electron joint momentum spectra}\label{secA2}

\begin{figure}
\centering
\includegraphics[width=1.0\textwidth]{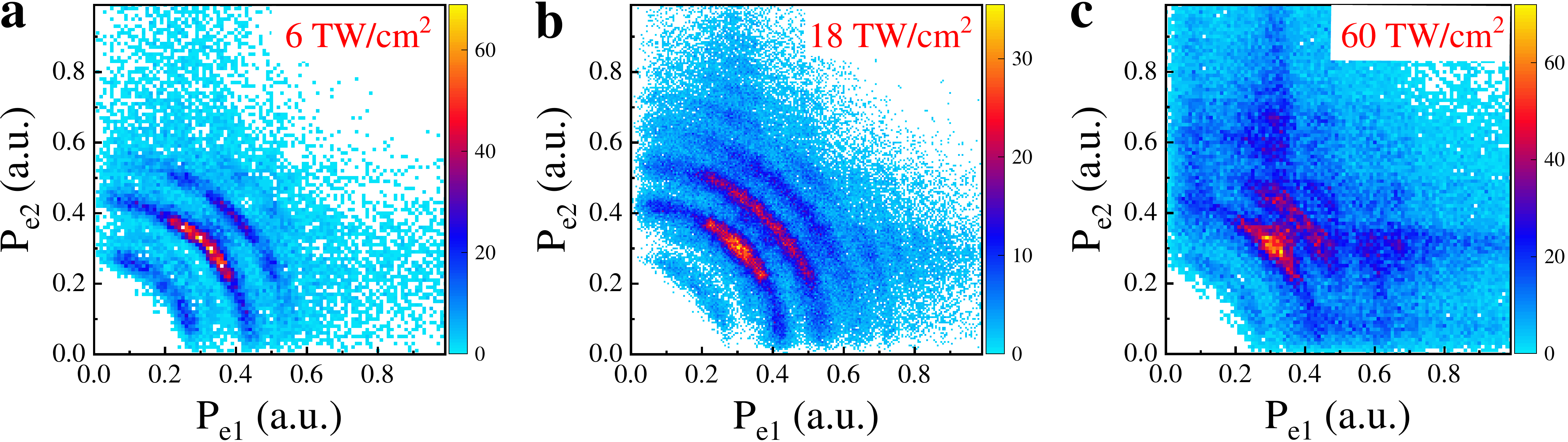}
\caption{Electron–electron JMS of Sr ($5s^2\ ^1S_0$) measured under linearly polarized 800~nm pulse at intensities of \textbf{a}, 6 TW/cm$^2$ , \textbf{b}, 18 TW/cm$^2$ and \textbf{c}, 60 TW/cm$^2$.}\label{Supplementary figure 2}
\end{figure}

Fig.~\ref{Supplementary figure 2} presents the electron-electron joint momentum spectra (JMS) of strontium atoms in the ground state irradiated at intensities of 6 TW/cm$^2$, 18 TW/cm$^2$, and 60 TW/cm$^2$. A series of circular arcs arise from the incoherent superposition of sequential and nonsequential ATDI. The strong correlation of electrons leading to nonsequential ATDI ($P_{e1} \approx P_{e2}$) dominates the DI process, consistent with the electron-electron joint energy spectra (JES) presented in Fig.~2a--c of the main text. 
For the linearly polarized pulse at an intensity of 6 TW/cm$^2$ ($U_p \approx 0.4 $~eV), the Keldysh parameter is calculated to be 2.7, indicating an ionization regime dominated by multi-photon absorption. The maximum energy of the recolliding electron is approximately $3.2 \times U_p \approx 1.3 $~eV, this energy is substantially below the second ionization potential of Sr (11.0~eV) and even lower than the energy of the first excited state of Sr$^+$ ($4p^64d$, 1.8~eV). Therefore, direct ionization (e, 2e) and recollision excitation followed by ionization can be excluded. Instead, electrons are populated in intermediate transition doubly excited states (DESs) by absorbing multiple photons, which leads to the stabilization of ATDI energy at elevated intensities.

\section{Details of theoretical methods}\label{secA3}

In the numerical simulation, the two-electron wave function is evolved under the following Hamiltonian
\begin{equation}
	H = -\frac{1}{2}\nabla_{1}^{2} - \frac{1}{2}\nabla_{2}^{2} + V(\bm{r}_1) + V(\bm{r}_2) - \ii \bm{A}(t)\cdot (\bm\nabla_1 + \bm\nabla_2) + \frac{1}{|\bm{r}_1 - \bm{r}_2|},
\end{equation}
where the laser field is
\begin{equation}
    \bm{A}(t) = A_0\hat{\bm{z}}\sin^2\frac{\pi t}{T}\cos\omega t,
	\label{eq: gaussian pulse}
\end{equation}
in which $T = 40\,\mathrm{fs}$ is the total length of the pulse, and $A_0$ is chosen to match a peak intensity of $I = 18\,\mathrm{TW}/\mathrm{cm}^2$.
The pseudopotential is~\cite{PFuentealba_1985}
\begin{equation}
    \label{eq: pspot}
    V(\bm{r}) = -\frac{Q}{r} + \sum_l \left\{B_l\exp(-\beta_lr^2) - \frac{\alpha_d}{2r^4}\left[1-\exp\left(-\frac{r^6}{r_{c,l}^6}\right)\right]\right\} P_l,
\end{equation}
in which $P_l$ is a projection operator onto the subspace of angular quantum number $l$, $Q = 2$ is the effective charge, and $\alpha_d = 5.813$ is the polarizability of the strontium core~\cite{Guo_2010}. The remaining $l$-dependent parameters are listed in Table \ref{Supplementary tab: pseudo}.
\begin{table}
  \centering
  \begin{tabular}{r|cccc}
    \hline
    $l$ & $0$ & $1$ & $2$ & $\ge3$\\\hline
    $B_l$ & $15.387$ & $5.077$ & $-2.248$ & $0$\\
    $\beta_l$ & $0.7962$ & $0.4212$ & $0.4927$ & $0$\\
    $r_{c,l}$ & $2.022$ & $1.950$ & $2.204$ & $1.761$\\
    \hline
  \end{tabular}
  \caption{Parameters of the pseudopotential.}
  \label{Supplementary tab: pseudo}
\end{table}

The basis set chosen for this calculation consists of the coupled spherical harmonics $Y_{l_1l_2}^{LM} = \sum_m\langle l_1\,l_2\,m\,(M-m)|L\,M\rangle$ for the angular part and a Gauss-Lobatto FE-DVR with non-uniform element sizes for the radial part. The angular momentum cutoff  is chosen to be $L_\text{max} = l_\text{max} = 7$ and the grid boundary  is $r_\text{max} = 250\,\mathrm{a.u.}$.

The initial state $5s^2$  is obtained with the Lanczos method, then time-propagated via the Krylov method. Through the course of propagation, a mask function  is periodically applied to prevent reflection. At each application, the split-off part consisting of amplitudes near the grid boundary  is projected into momentum space and evolved subsequently according to a simple Volkov phase: $\exp\{-\ii[k_1^2/2+k_2^2/2 + \bm{A}\cdot(\bm{k_1}+\bm{k_2})]t\}$.

The final wave function  is projected into the momentum space and then combined with the coherent sum of all previous split-off parts to obtain the momentum distribution $P(\bm{k}_1, \bm{k}_2)$, which can be  further integrated over angles for the JES. 

\subsection{Absorption of outgoing waves}\label{secA2_1}
Through the course of propagation, a mask function in the form of $f(r_1)\cdot f(r_2)$ is applied periodically, with  $f(r)$ given by:
\begin{equation}
    \label{eq: mask}
    f(r) = \begin{cases}
        1, & r < r_c,\\
        \displaystyle 1-\left(\frac{r-r_c}{r_d}\right)^2\cdot\left(3 - 2\cdot\frac{r-r_c}{r_d}\right), & r_c\le r \le r_c+r_d,\\
        0, & r > r_c+r_d,
    \end{cases}
\end{equation}
with parameters $r_c = 100\,\mathrm{a.u.}$ and $r_d = 100$~a.u. This mask function is chosen to preserve the continuity of the derivative at both $r_c$ and $r_c + r_d$ to provides a smooth splitting.  


\subsection{Momentum space projection}\label{secA2_2}
The basis of the momentum space are approximate two-electron scattering states, taken to be the symmetrized product of two independent single particle scattering states:
\begin{equation}
    \phi_{\bm{k}_1\bm{k}_2}(\bm{r}_1,\bm{r}_2) = \frac{1}{\sqrt{2}}\left[\phi_{\bm{k}_1}(\bm{r}_1)\phi_{\bm{k}_2}(\bm{r}_2) + \phi_{\bm{k}_1}(\bm{r}_2)\phi_{\bm{k}_2}(\bm{r}_1)\right].
\end{equation}
The single-particle scattering states are the ingoing solutions of the single-electron Schr\"odinger equation, with the form
\begin{equation}
    \phi_{\bm{k}}(\bm{r}) = \sum_{l=0}(2l+1)\ii^l\ee^{-\ii\delta_l}R_{kl}(r)P_l(\cos\theta),
\end{equation}
where $\delta_l$ is the Coulomb phase shift to be determined.

The $l$-separated radial solutions $R_{kl}$ are obtained by solving the radial equation
\begin{equation}
    \label{eq: radial}
    \left[-\frac{1}{2}\frac{\dd{{}^2}}{\dd{r}^2} - \frac{1}{r}\frac{\dd{}}{\dd{r}} + \frac{l(l+1)}{2r^2} + V_l(r) - \frac{1}{2}k^2\right] R_{kl}(r) = 0,
\end{equation}
with the Killingbeck-Miller method~\cite{JPKillingbeck_1985} with sufficiently small radial step $h$. To determine the phase shift $\delta_l$ and normalization constant $a$, the unnormalized $\tilde{R}_{kl}(r)$  is compared against the free space solution with phase shift in the asymptotic region:
\begin{equation}
    R_{kl}(r) = a\cdot \tilde{R}_{kl}(r) \xrightarrow[r\to\infty]{} \frac{\sin(kr - l\pi/2 + \delta_l)}{kr}.
\end{equation}
In practice, this is done by comparing the position and magnitude of any single extremum of $kr\tilde{R}_{kl}(r)$ against the sine function at $kr\approx 1000$.




\bibliography{sn-bibliography}